\documentclass[11pt]{article}
\usepackage[utf8]{inputenc}
\usepackage{mathtools,amsmath,amssymb}
\usepackage{graphicx}
\usepackage{lmodern}
\usepackage[T1]{fontenc}
\usepackage{microtype}
\usepackage[pagebackref=false]{hyperref}
\numberwithin{equation}{section}

\usepackage[textsize=tiny]{todonotes}

\usepackage{xspace}
\usepackage{bibentry}
\usepackage{easybmat}

\usepackage{graphicx}

\usepackage{subcaption} 

\usepackage[numbers,sort&compress]{natbib}
\usepackage{hypernat}
 
\usepackage{tikz}
\usetikzlibrary{decorations.pathreplacing}
 
\newcommand{\changelocaltocdepth}[1]{%
  \addtocontents{toc}{\protect\setcounter{tocdepth}{#1}}%
  \setcounter{tocdepth}{#1}%
}

\usepackage[margin=2.5cm]{geometry} 
    
\newcommand{\ID}{I}

\newmuskip\pFqmuskip

\newcommand*\pFq[6][8]{%
  \begingroup 
  \pFqmuskip=#1mu\relax
  \mathchardef\normalcomma=\mathcode`,
  \mathcode`\,=\string"8000
  \begingroup\lccode`\~=`\,
  \lowercase{\endgroup\let~}\pFqcomma
  {}_{#2}\phi_{#3}{\left[\genfrac..{0pt}{}{#4}{#5};#6\right]}%
  \endgroup
}
\newcommand{\pFqcomma}{{\normalcomma}\mskip\pFqmuskip}

\DeclareMathOperator{\tr}{tr}

  \newcommand{\ptwo}{
\begin{tikzpicture}[scale=0.4] 
    \fill[black!30] (5,0) rectangle +(1,1);
    \fill[black!30] (4,1) rectangle +(1,1);
    \fill[black!30] (3,2) rectangle +(1,1);
    \fill[black!30] (2,3) rectangle +(1,1);
    \fill[black!30] (1,4) rectangle +(1,1);
    \fill[black!30] (0,5) rectangle +(1,1);

    \fill[red!30] (3,4) rectangle +(1,1); 
    \fill[red!30] (3,3) rectangle +(1,1); 
   \fill[red!50] (3,2) rectangle +(1,1); 
      \fill[red!50] (2,3) rectangle +(1,1); 
         \fill[red!30] (4,3) rectangle +(1,1); 
    \draw[black, very thin] (0,0) grid (6,6);
\end{tikzpicture}}
  \newcommand{\pone}{
\begin{tikzpicture}[scale=0.4] 
    \fill[black!30] (5,0) rectangle +(1,1);
    \fill[black!30] (4,1) rectangle +(1,1);
    \fill[black!30] (3,2) rectangle +(1,1);
    \fill[black!30] (2,3) rectangle +(1,1);
    \fill[black!30] (1,4) rectangle +(1,1);
    \fill[black!30] (0,5) rectangle +(1,1);

    \fill[red!30] (4,4) rectangle +(1,1); 
    \fill[red!30] (4,3) rectangle +(1,1); 
   \fill[red!30] (4,2) rectangle +(1,1); 
      \fill[red!30] (3,3) rectangle +(1,1); 
         \fill[red!30] (5,3) rectangle +(1,1); 
    \draw[black, very thin] (0,0) grid (6,6);
\end{tikzpicture}}
  \newcommand{\pthree}{
\begin{tikzpicture}[scale=0.4] 
    \fill[black!30] (5,0) rectangle +(1,1);
    \fill[black!30] (4,1) rectangle +(1,1);
    \fill[black!30] (3,2) rectangle +(1,1);
    \fill[black!30] (2,3) rectangle +(1,1);
    \fill[black!30] (1,4) rectangle +(1,1);
    \fill[black!30] (0,5) rectangle +(1,1);

    \fill[red!30] (2,4) rectangle +(1,1); 
    \fill[red!50] (2,3) rectangle +(1,1); 
   \fill[red!30] (2,2) rectangle +(1,1); 
      \fill[red!30] (1,3) rectangle +(1,1); 
         \fill[red!30] (3,3) rectangle +(1,1); 
    \draw[black, very thin] (0,0) grid (6,6);
\end{tikzpicture}}

 \newcommand{\pfour}{
\begin{tikzpicture}[scale=0.4] 
    \fill[black!30] (5,0) rectangle +(1,1);
    \fill[black!30] (4,1) rectangle +(1,1);
    \fill[black!30] (3,2) rectangle +(1,1);
    \fill[black!30] (2,3) rectangle +(1,1);
    \fill[black!30] (1,4) rectangle +(1,1);
    \fill[black!30] (0,5) rectangle +(1,1);

    \fill[red!50] (2,3) rectangle +(1,1); 
    \fill[red!30] (2,2) rectangle +(1,1); 
   \fill[red!30] (2,1) rectangle +(1,1); 
      \fill[red!30] (1,2) rectangle +(1,1); 
         \fill[red!50] (3,2) rectangle +(1,1); 
    \draw[black, very thin] (0,0) grid (6,6);
\end{tikzpicture}}
 
\newcommand{\pfive}{
\begin{tikzpicture}[scale=0.4] 
    \fill[black!30] (5,0) rectangle +(1,1);
    \fill[black!30] (4,1) rectangle +(1,1);
    \fill[black!30] (3,2) rectangle +(1,1);
    \fill[black!30] (2,3) rectangle +(1,1);
    \fill[black!30] (1,4) rectangle +(1,1);
    \fill[black!30] (0,5) rectangle +(1,1);

    \fill[red!30] (2,2) rectangle +(1,1); 
    \fill[red!30] (2,1) rectangle +(1,1); 
   \fill[red!30] (2,0) rectangle +(1,1); 
      \fill[red!30] (1,1) rectangle +(1,1); 
         \fill[red!30] (3,1) rectangle +(1,1); 
    \draw[black, very thin] (0,0) grid (6,6);
\end{tikzpicture}}

\begin{document} 
 
\begingroup
\begin{center}
 \begingroup\LARGE
\bf Integrable boundaries for the q-Hahn process
\par\endgroup
 \vspace{3.5em}
 \begingroup\large \bf
 {Rouven Frassek}
 \par\endgroup
\vspace{2em}

\begingroup\sffamily 
%
University of Modena and Reggio Emilia, 
\\Department of Physics, Informatics and Mathematics,\\
Via G. Campi 213/b, 41125 Modena, Italy\\ 
\par\endgroup
\vspace{2em}

\end{center}

\thispagestyle{empty}

\begin{abstract}
\noindent
Taking inspiration from the  harmonic process with reservoirs introduced by Giardin\`{a}, Kurchan and the author in \cite{Frassek:2019vjt}, we propose  integrable boundary conditions for its trigonometric deformation, which is known as the  q-Hahn process. Following the formalism established by Mangazeev and Lu in  \cite{Mangazeev:2019rzf} using the stochastic R-matrix,
 we  argue that the proposed boundary conditions can be derived from a transfer matrix constructed in the framework of Sklyanin's extension of the quantum inverse scattering method and consequently preserve the integrable structure of the model. The  approach avoids the explicit construction of the K-matrix.
\end{abstract}

\vfill 

\tableofcontents


\section{Introduction} 
The symmetric and asymmetric simple exclusion processes have received a lot of attention in the last decades. One of the reasons is that they are exactly solvable and thus exhibit a profound underlying mathematical structure. A particularly interesting setup that preserves the integrability is when the SSEP/ASEP is in contact with two reservoirs at the left and right end of the chain. In this setup the model is still integrable but more difficult to study, e.g. the steady state does not factorise any longer. The processes are known to be related to the XXX and the XXZ Heisenberg spin chain and thus can be formulated in the framework of the quantum inverse scattering method \cite{Faddeev:1996iy} that was generalised  in \cite{Sklyanin:1988yz} to the case of open boundaries which are relevant to model the reservoirs of the processes. 

More recently an integrable generalisation of the ASEP was proposed that is known as the q-Hahn process. In a certain regime of parameters this model has already appeared two decades ago \cite{Sasamoto} but the full jump probabilities  were only recently obtained by Povolotsky \cite{Povolotsky}. The rates and the extension to right and left particle jumps was then given in \cite{barraquand}.
  Similar as in the case of the ordinary ASEP, the generator of the q-Hahn process is related to the  Hamiltonian of the $U_q(sl_2)$ invariant non-compact  XXZ spin chain with a certain choice of parameters by a similarity transformation as discussed in \cite{Frassek:2019isa}. The relevant representations on the infinite-dimensional vector space $V$ whose tensor product forms the  quantum space of the spin chain is defined by the following action of the generators
\begin{equation}\label{eq:sl2act}
 S_+|m\rangle=[m+2s]|m+1\rangle\,,\quad S_-|m\rangle=[m]|m-1\rangle\,,\quad S_0|m\rangle=(m+s)|m\rangle\,.
\end{equation}  
Here the spin label is a positive real number, i.e. $s>0$. The  generators obey the standard $U_q(sl_2)$ commutation relations
\begin{equation}\label{eq:sl2}
 [S_+,S_-]=-[2S_0]\,,\quad\quad [S_0,S_\pm]=\pm S_\pm\,,
\end{equation} 
with the q-number defined via
\begin{equation}
 [x]=\frac{q^{x}-q^{-x}}{q-q^{-1}}\,.
\end{equation}  
The algebraic formulation is advantageous in particular when studying stochastic dualities of the processes see e.g. \cite{Schtz1994NonAbelianSO,giardina2009duality} but not to describe the actual process which for a given representation arises in a specific choice of basis.

The main goal of this note is to introduce boundary reservoirs for the continuous time q-Hahn process that preserve the integrable structure. This has been done in \cite{Frassek:2019vjt} for the rational $q\to1$ limit following the  quantum inverse scattering method. The authors derived a suitable K-matrix written in terms of $sl_2$ generators (similar to the evaluation map in the case of the ordinary Yangian) and were able to show that it yields the desired boundary terms following Sklyanin's prescription \cite{Sklyanin:1988yz} by computing the relevant matrix elements of the resulting boundary Hamiltonian. It should be possible to apply the same strategy for the q-deformed case but to our knowledge we are currently lacking a suitable expression of the most general K-matrix. To circumvent the construction of the K-matrix we follow an idea initially proposed by Sutherland  \cite{sutherland1970two} for the Hamiltonian density and generalise it to our setup with boundaries. 
For this purpose  it is convenient to fix the basis and work right from the beginning in components. We thus follow closely the conventions of  \cite{Mangazeev:2019rzf}  to avoid algebraic expressions.

  The note is organised as follows: Section~\ref{sec:qhahn} summarises our results and presents  the Markov generator of  the open q-Hahn process. Section~\ref{sec:manga} reviews the formalism and results of \cite{Mangazeev:2019rzf} that are needed in order to argue that the proposed boundary conditions are integrable. This latter point is discussed in Section~\ref{sec:ham}. The argument is based on solving the boundary Yang-Baxter equation and its first derivative with respect to the spectral parameter at special points that are relevant in the construction of the spin chain Hamiltonian. Section~\ref{sec:otasep} briefly discusses the open q-TASEP limit and Section~\ref{sec:conc} contains our conclusions. Further details of computations can be found in the appendix.

\section{The open q-Hahn process}\label{sec:qhahn}
In this section we propose integrable boundary conditions for the q-Hahn process.  The full stochastic Hamiltonian of the open non-compact XXZ spin chain with $N$ sites and reservoirs at site $1$ and $N$ is of the form
\begin{equation}\label{eq:ham}
H=B_L+\sum_{i=1}^{N-1}\mathcal{H}_{i,i+1}+B_R
\end{equation} 
where 
\begin{equation}\label{eq:qspace} 
 H: V\otimes \ldots \otimes V \rightarrow V\otimes \ldots \otimes V\,.
\end{equation} 
Here we introduced the Hamiltonian density $\mathcal{H}_{i,i+1}$ that acts non-trivially on the $i$th and $(i+1)$th space along with the boundary terms $B_L$ and $B_R$ that act non-trivially on the first and last site of the spin chain respectively. The explicit form of the Hamiltonian density and the boundary terms will be given in the following subsections.  The given  Hamiltonian then  relates to the continuous-time Markov generator $M$ with standard conventions via the simple relation
\begin{equation}
 M=-H^{t}\,,
\end{equation} 
where $t$ denotes the transposition in the quantum space $V\otimes \ldots\otimes V$.
Further, as can be checked by an explicit computation, the introduced  Hamiltonian  is stochastic since all matrix elements outside
the diagonal are negative or zero. The sum over the columns vanishes. 
The bulk part has been introduced in \cite{Sasamoto,Povolotsky,barraquand}, see also \cite{Frassek:2019isa}, and is discussed in Section~\ref{sec:bulkH}. The boundary part is given in Section~\ref{sec:bnd}. Each boundary term introduced  at site $1$ and $N$ is governed by one additional  parameter $\rho_L$ and $\rho_R$ respectively.
\subsection{Bulk}\label{sec:bulkH}
The  q-Hahn process in the bulk is obtained from the stochastic Hamiltonian density defined 
by  the action on two neighboring sites with  $m$ particles on one site and $m'$ particles on the other via
\begin{equation}\label{eq:hact}
\begin{split}
\mathcal{H}|m\rangle \otimes |m'\rangle=\left(\alpha_-(m)+\alpha_+(m')\right)|m\rangle \otimes |m'\rangle&-\sum_{k=1}^m \beta_-(m,k) |m-k\rangle \otimes |m'+k\rangle\\&-\sum_{k=1}^{m'} \beta_+(m',k) |m+k\rangle \otimes |m'-k\rangle\,.
 \end{split}
\end{equation} 
The rates for particles moving to the left are of the form
\begin{equation}\label{eq:bp}
\beta_+(m,k)= 
\frac{\mu ^{k} }{1-\gamma^k}\frac{(\gamma;\gamma)_m (\mu ;\gamma)_{m-k}}{ (\gamma;\gamma)_{m-k} (\mu ;\gamma)_m}\,,
\end{equation} 
while the rates for particles that move to the right read
\begin{equation}\label{eq:bm}
\beta_-(m,k)= 
\frac{1}{1-\gamma^k}\frac{(\gamma;\gamma)_m (\mu ;\gamma)_{m-k}}{ (\gamma;\gamma)_{m-k} (\mu ;\gamma)_m}\,.
\end{equation}  
Here the parameters take values $0<\gamma,\mu<1$ and  $(a;\gamma)_m=\prod_{j=0}^{m-1}(1-a\gamma^j)$ denotes
the q-Pochhammer symbol.
The diagonal terms are expressed in terms of finite sums as
\begin{equation}\label{eq:alphas}
\alpha_+(m)
=\sum _{k=0}^{m-1} \frac{\gamma^{k}}{\mu^{-1}-\gamma^{k}}\,,\qquad 
 \alpha_-(m)
 =\sum _{k=0}^{m-1} \frac{1}{1-\gamma^{k}\mu}\,,
\end{equation} 
see also  Appendix~\ref{app:sumrates}.
For later purposes we set 
\begin{equation}\label{eq:qq}
 \gamma=q^2\,,\qquad \mu=q^{4s}\,,
\end{equation}  
where $q$ is the deformation parameter of the universal enveloping algebra $U_q(sl_2)$ in \eqref{eq:sl2}.
\subsection{Boundary}\label{sec:bnd}
Taking inspiration from \cite{Frassek:2019vjt}, we introduce the boundary terms 
\begin{equation}\label{eq:bd1}
\begin{split}
B_L|m\rangle =\left(\alpha_+(m) +\sum_{k=1}^\infty\frac{\rho_L^k}{1-\gamma^{k}}\right)|m\rangle&
-\sum_{k=1}^m \beta_+(m,k)|m-k\rangle -\sum_{k=1}^\infty\frac{\rho_L^k}{1-\gamma^{k}}|m+k\rangle \,,
\end{split}
\end{equation} 
and
\begin{equation}\label{eq:bd2}
\begin{split}
B_R|m\rangle =\left(\alpha_-(m) +\sum_{k=1}^\infty\frac{\rho_R^k}{1-\gamma^{k}}\right)|m\rangle&
-\sum_{k=1}^m\beta_-(m,k)|m-k\rangle -\sum_{k=1}^\infty\frac{\rho_R^k}{1-\gamma^{k}}|m+k\rangle \,.
\end{split}
\end{equation} 
Particles are removed from the first and last site with the  bulk rates into the reservoirs. The insertion rates are governed by one real parameter  at each boundary $\rho_L$ and $\rho_R$ with $0<\rho_L,\rho_R<1$ as in the rational case.
We are free to shift $\rho_R\to\rho_R\mu$ in order to achieve an expression more similar to the bulk  rates in \eqref{eq:bp} and \eqref{eq:bm}.

\section{Quantum inverse scattering method}\label{sec:manga}

In the framework of the quantum inverse scattering method the Hamiltonian is derived from the fundamental transfer matrix. More precisely it is obtained from the logarithmic derivative at the permutation point $x_0$, i.e.
\begin{equation}\label{eq:logderiv1}
H=-\frac{1}{4}\frac{\partial}{\partial x}\ln T(x)\big|_{x=x_0} +c_0\,\ID \,,
\end{equation} 
with a constant $c_0$ multiplying the identity and which we will fix later on.
The transfer matrix is given by Sklyanin's double-row construction \cite{Sklyanin:1988yz}   and is expressed in terms of R- and K-matrices  via
\begin{equation}\label{eq:transm}
 T(x)=\tr_a \bar{\mathcal{K}}_a(x)\mathcal{R}_{a,1}(x)\cdots \mathcal{R}_{a,N}(x)\mathcal{K}_a(x)\mathcal{R}_{N,a}(x^{-1})\cdots \mathcal{R}_{1,a}(x^{-1})\,.
\end{equation} 
Here we  use the notation
\begin{equation}
 \mathcal{R}(x)= \mathcal{R}_{a,i}(x)\,,
\end{equation} 
and 
\begin{equation}\label{eq:symm}
 \mathcal{R}_{i,a}(x)=\mathcal{P}\,\mathcal{R}_{a,i}(x)\,\mathcal{P}=\mathcal{R}^{-1}(x^{-1})\,,
\end{equation} 
where $i=1,\ldots, N$ denote the spin chain sites and $\mathcal{P}$ the permutation operator.  The R-matrices act non-trivially on the tensor product of two vector spaces $V\otimes V$ as introduced above   in \eqref{eq:qspace}   while the K-matrices act non-trivially on a single copy of the auxiliary space $V_a\sim V$ isomorphic to a single site of the quantum space.

Additionally we introduce the transfer matrix with the two-dimensional defining representation in the auxiliary space 
\begin{equation}\label{eq:bat}
 T_\square(x)=\tr_\square  \bar{K}_\square (x)\mathcal{L}_{\square,1}(x)\cdots \mathcal{L}_{\square ,N}(x){K}_\square (x)\mathcal{L}_{N,\square }(x^{-1})\cdots \mathcal{L}_{1,\square }(x^{-1})\,,
\end{equation} 
here indicated with $\square$.
The Lax matrices $\mathcal{L}$ act non-trivially on $\mathbb{C}^2$ in the auxiliary space and the quantum space at a single site $V$. The K-matrices act non-trivially on the auxiliary space and are matrices of the size $2\times2$. This transfer matrix is usually used in the context of the algebraic Bethe ansatz \cite{Sklyanin:1988yz}. 

To ensure commutativity of the transfer matrix the boundary Yang-Baxter relations 
\begin{equation}\label{eq:bybe1}
 \mathcal{L}_{a,\square }(x/y)\mathcal{K}_a(x) \mathcal{L}_{\square,a}(xy)K_\square(y)=K_\square(y) \mathcal{L}_{a,\square}(xy)\mathcal{K}_a(x)  \mathcal{L}_{\square,a}(x/y)
\end{equation} 
and
\begin{equation}\label{eq:bybe2}
  \mathcal{L}_{a,\square}(y/x)\bar{\mathcal{ K}}_a(x) \bar{ \mathcal{L}}_{\square,a}(1/(xy))\bar K_\square(y)=\bar K_\square(y)\bar{ \mathcal{L}}_{a,\square}(1/(xy))\bar {\mathcal{K}}_a(x) { \mathcal{L}}_{\square,a}(y/x)
\end{equation} 
for the left and right boundary have to be obeyed. For a detailed discussion that also takes into account the bulk part we refer the reader to \cite{Mangazeev:2019rzf}. 

The Lax matrices $\bar{\mathcal{L}}$ in the second boundary Yang-Baxter relation \eqref{eq:bybe2}  is defined via
\begin{equation}\label{eq:Lbars}
 \bar{ \mathcal{L}}_{\square, a}(x)=\left(\left(\mathcal{L}_{a,\square}^{t_\square}(1/x)\right)^{-1}\right)^{t_\square}\,,\qquad \bar{ \mathcal{L}}_{a,\square}(x)=\left(\left(\mathcal{L}_{\square ,a}^{t_a}(1/x)\right)^{-1}\right)^{t_a} \,.
\end{equation} 
We will express them in terms of the Lax matrices $\mathcal{L}$ in Section~\ref{sec:cros} using the appropriate crossing relations.
The relevant R- and Lax matrices  can be found in the component form in \cite{Mangazeev:2019rzf}.  
The Lax matrices read
\begin{equation}
\langle m|\mathcal{L}_{\square ,a}(x)|m'\rangle=\frac{1}{ \left[x q^{\frac{1}{2}-s}\right]}\left(\begin{array}{cc}
                      \delta_{m,m'}\left[x q^{\frac{1}{2}-s-m}\right]q^{m}&\delta_{m,m'+1}x\left[q^{1-2s-m}\right] q^{m-\frac{1}{2}-s}\\\delta_{m,m'-1}
                      x^{-1}\left[q^{m+1}\right] q^{\frac{1}{2}+s+m}&\delta_{m,m'}\left[x q^{m+s+\frac{1}{2}}\right]q^{m+2s}
                     \end{array}
\right)
\end{equation} 
and
\begin{equation}
\langle m|\mathcal{L}_{a,\square}(x)|m'\rangle=\frac{1}{ \left[x q^{\frac{1}{2}-s}\right]}\left(\begin{array}{cc}
                      \delta_{m,m'}\left[x q^{\frac{1}{2}-s-m}\right]q^{-m}&\delta_{m,m'+1}x^{-1}\left[q^{1-2s-m}\right] q^{-m-s+\frac{1}{2}}\\\delta_{m,m'-1}
                      x\left[q^{m+1}\right] q^{-\frac{1}{2}-s-m}&\delta_{m,m'}\left[x q^{m+s+\frac{1}{2}}\right]q^{-m-2s}
                     \end{array}
\right)\,.
\end{equation} 
They obey the unitarity relations $\mathcal{L}_{a,\square}(x)
 \mathcal{L}_{\square,a}(x^{-1})=\ID$. 
The R-matrix for the fundamental transfer matrix \eqref{eq:transm} is expressed through its components  in the same article \cite{Mangazeev:2019rzf}. It is written in terms of a basic hypergeometric function, cf.~\cite{gasper2004basic}, as
\begin{equation}\label{eq:Rmatrix1}
\begin{split}
 \langle m,n|\mathcal{R}(x)|m',n'\rangle =\delta_{m+n,m'+n'}q^{4ms} &
 \left[\begin{array}{c}
 m+n                                                                      \\
m                                                                           \end{array}
\right]_{q^2}
\frac{\left(x^{-2};q^{2}\right)_{n'}\left(x^{-2};q^2\right)_{m}\left(q^{4s};q^2\right)_{n}}{\left(x^{-2}q^{4s};q^2\right)_{m+n}\left(q^{4s};q^2\right)_{n'}}\\
&\hspace{-1.15cm}\times\,\,\,\pFq{4}{3}{ q^{-2m},q^{-2n'},x^2 q^{4s},x^2 q^{2-4s-2m-2n}}{q^{-2m-2n},x q^{2-2m},x q^{2-2n'}}{q^2,q^2}\,.
\end{split}
\end{equation} 
where the q-binomial is defined by
\begin{equation}
  \left[\begin{array}{c}
 n                                                                      \\
m                                                                           \end{array}
\right]_{q}=\frac{(q;q)_n}{(q;q)_{n-m}(q;q)_m}\,.
\end{equation} 
The  R-matrix is not symmetric under the exchange $m\leftrightarrow n$ and $m'\leftrightarrow n'$. 
We further remark that there is a diagonal Drinfeld twist involved \cite{borodin2016stochastic},  see also \cite[(3.12)]{Mangazeev:2019rzf}.  This is not immediately relevant for us but convenient as the logarithmic derivative will immediately yield a Hamiltonian on the stochastic line. The twist can also be introduced  at a later stage and appears in the form of a similarity transformation on the level of the Hamiltonian, cf.~\cite{deGier:2005zz,Frassek:2019isa}.

For later purposes we also introduce an  alternative expression  for the R-matrix. 
In  \cite{Mangazeev:2019rzf} it has been written as
\begin{equation}\label{eq:Rmatrix2}
\begin{split}
 \langle m,n|\mathcal{R}_{a,i}(x)|m',n'\rangle =\delta_{m+n,m'+n'} \sum_{k=0}^{m}\Phi_{q^2}(k|k+n;x^{-2},x^{-2}q^{4s})\Phi_{q^2}(m-k|n';x^2q^{4s},q^{4s})
\end{split}
\end{equation} 
with 
\begin{equation}\label{eq:phii}
 \Phi_{q}(m|n;x,y)=\left(\frac{y}{x}\right)^m
\frac{\left(x;q\right)_{m}\left(y/x;q\right)_{n-m} }{\left(y;q\right)_{n }}
 \left[\begin{array}{c}
 n                                                              \\
m                                                                    \end{array}
\right]_{q}\,.
\end{equation} 
This factorisation reminds of Derkachov's factorisation in relation to Q-operators \cite{Derkachov:1999pz}. We will see later in Section~\ref{sec:bulkT} that the product of the two terms  turns into a sum at the level of the Hamiltonian density. One term yields the left moving particles and the other the right moving particles which combine into the q-Hahn process.  

Finally we also introduce the K-matrices in the two-dimensional defining representation that appears in the definition of the transfer matrix \eqref{eq:bat}. It depends on the four additional free parameters $t^\pm_R,\nu_R,\kappa_R$ that will be adjusted later and reads
\begin{equation}
 K_\square(x)=\left(\begin{array}{cc}
                     \frac{t^-_R}{q\nu_R}-q\nu_R t^+_R+x^2(t^-_R-t^+_R)&\kappa_R^{-1}t_R^-(x^2-x^{-2})\\
                     \kappa_R t_R^+(x^2-x^{-2})& \frac{t_R^-}{q\nu_R}-q\nu_R t_R^++x^{-2}(t_R^--t_R^+)
                    \end{array}
\right)\,.
\end{equation} 
\cite{Vega,GZ:2005zz}.
As discussed in   \cite{Mangazeev:2019rzf}  we can express the other K-matrix via
\begin{equation}\label{eq:crossK}
 \bar K_\square(x)=\left.D_\square^{-1}K_\square(1/(qx))\right|_{R\to L}\,,
\end{equation} 
where we introduced another four variables $t^\pm_L,\nu_L,\kappa_L$ and the matrix 
\begin{equation}
 D_\square=\left(\begin{array}{cc}
                                                         1&0\\
                                                         0&q^2
                                                        \end{array}
\right) \,.
\end{equation} 
To our knowledge, the K-matrices $\mathcal{K}$ and $\bar{\mathcal{K}}$ in \eqref{eq:bybe1} and \eqref{eq:bybe2} seem currently not available in the literature in the infinite-dimensional representations of interest.

\subsection{Crossing relations and the other boundary}\label{sec:cros}
In this subsection we discuss the crossing relations for the Lax matrices introduced above and argue that the boundary Yang-Baxter equation \eqref{eq:bybe2} can be brought to the form \eqref{eq:bybe1}. From the latter statement it then follows that the left and right K-matrices are determined by the same equation \eqref{eq:bybe1}. 
The crossing relations have been given in \cite{Mangazeev:2019rzf}. The first one reads
\begin{equation} 
\left(
 \mathcal{L}_{s,\square}^{t_\square}(1/ x)\right)^{-1}=g(x) D_\square  \mathcal{L}_{\square,s}^{t_\square}(x q^{-2})D_\square^{-1}
\end{equation} 
with 
\begin{equation}
 g(x)=\frac{\left(q^{s+3}-x^2\right) \left(x^2 q^s-q\right)}{\left(q^{s+1}-x^2\right) \left(x^2 q^s-q^3\right)}\,.
\end{equation}  
It can be used  to express  the Lax matrix appearing in \eqref{eq:bybe2} as
\begin{equation}
 \bar{\mathcal{L}}_{\square,s}(x)=g(x)D_\square^{-1}\mathcal{L}_{\square,s} (x q^{-2})D_\square\,.
\end{equation}  
The second crossing relation relevant is of the form
\begin{equation} 
\left(
 \mathcal{L}_{\square,s}^{t_s}(1/x)\right)^{-1}=g(x) D_s\mathcal{L}_{s,\square}^{t_s}(x q^{-2})D_s^{-1}
\end{equation} 
with the same function $g(x)$. The similarity transformation is diagonal and  of the form
\begin{equation}\label{eq:dmat}
 \langle m|D_s|m'\rangle =q^{2m}\delta_{m,m'} \,.
\end{equation} 
Similar as before we obtain
\begin{equation}
\bar{ \mathcal{L}}_{s,\square}(x)=g(x)D_s^{-1}S_{s,\square}(x q^{-2})D_s\,,
\end{equation} 
cf.~\eqref{eq:bybe2}. 
We insert the obtained relations into the boundary Yang-Baxter equation \eqref{eq:bybe2}. Further,  inverting the K-matrix $K_\square$ and $D_\square$ and shifting the spectral parameters $y\to y q^{-1}$ and $x\to (xq)^{-1}$ we obtain
\begin{equation}
   \bar K_\square^{-1}(y/q)D_\square^{-1} S_{s,\square}(y x)D_s\bar K_s(1/(xq)) {S}_{\square,s}(x/y)={S}_{s,\square}(x/y)D_s\bar K_s(1/(xq)) S_{\square,s}(xy)\bar K_\square^{-1}(y/q)D_\square^{-1}\,.
\end{equation} 
Here we also used the invariance condition $
 [D_s D_\square,S_{s,\square}(y x)]=0$.
As  the K-matrix obeys the relation \eqref{eq:crossK}, it follows that the inverse of $\bar K_\square$ can be expressed as
\begin{equation}
 \bar K_\square^{-1}(y/q)D_\square^{-1} =\left.K_\square^{-1} (1/y)\right|_{R\to L}\,.
\end{equation} 
Further it is easy to verify that 
\begin{equation}
 K_\square^{-1} (1/y)=\frac{q^2 \nu_R^2 y^4}{\left(q \nu_R+y^2\right) \left(q \nu_R y^2+1\right) \left(q t^+_R\nu_R-t^-_R y^2\right) \left(q t^+_R\nu_R y^2-t^-_R\right)} K_\square (y)\,.
\end{equation} 
Inserting also this relation in the second boundary Yang-Baxter equation we find that solutions to the two boundary Yang-Baxter equations \eqref{eq:bybe1} and \eqref{eq:bybe2} can be related via
\begin{equation}
 \bar{\mathcal{K}}_s(x)=\left.D_s^{-1}\mathcal{K}_s(1/(qx))\right|_{R\to L}\,.
\end{equation} 
This relation will be of importance in Section~\ref{sec:ham}.

\subsection{Component form of BYBE}
We end this  section with the boundary Yang-Baxter equation \eqref{eq:bybe1} in component form as given in \cite{Mangazeev:2019rzf}. The equations below can be obtained taking coefficients in spectral parameter $x$:
\begin{equation}\label{manga1}
 \begin{split}
 \kappa_R t^+_R q^{2-4s}\left(1-q^{2(j+2s)}\right)\mathcal{K}_{j,l+1}(y)
 +\kappa^{-1}_Rt^-_Rq^{-4s}\left(1-q^{2+2l}\right)\mathcal{K}_{j+1,l}(y)\\
 +\nu^{-1}_Ry^2q^{2-2s}\left(q^{2j}-q^{2l}\right)\left(t^-_R-\nu^2_Rq^2t^+_R\right)\mathcal{K}_{j+1,l+1}(y)
 -\kappa_R t^+_R y^4q^{2-4s}\left(1-q^{2(1+l+2s)}\right)\mathcal{K}_{j+1,l+2}(y)\\
 -\kappa^{-1}_Rt^-_Ry^4 q^{-4s}\left(1-q^{4+2j}\right)\mathcal{K}_{j+2,l+1}(y)=0\,,
 \end{split}
\end{equation} 
and
\begin{equation}\label{manga2}
 \begin{split}
 \kappa_R t^+_R q^{2(2+l-2s)}\left(1-q^{2(j+2s)}\right)\mathcal{K}_{j,l+1}(y)
 +\kappa^{-1}_Rt^-_Rq^{2(1+j-2s)}\left(1-q^{2+2l}\right)\mathcal{K}_{j+1,l}(y)\\
 +q^{2-4s}\left(q^{2j}-q^{2l}\right)\left(t^+_R-t^-_R\right)\mathcal{K}_{j+1,l+1}(y)
 -\kappa_R t^+_R q^{2(1+j-2s)}\left(1-q^{2(1+l+2s)}\right)\mathcal{K}_{j+1,l+2}(y)\\
 -\kappa^{-1}_Rt^-_R q^{-4s+2l}\left(1-q^{4+2j}\right)\mathcal{K}_{j+2,l+1}(y)=0\,.
 \end{split}
\end{equation} 
Here we use the notation $\mathcal{K}_{i,j}(x)=\langle i|\mathcal{K}(x)|j\rangle $. The equations above yield conditions on five neighboring entries of the K-matrix.  This is discussed in further detail in Section~\ref{sec:rib}.

\section{Stochastic Hamiltonian from the transfer matrix}\label{sec:ham}

Let us compute the logarithmic derivative of the fundamental transfer matrix \eqref{eq:transm}.   
Assuming that the K-matrix at $x=1$ is proportional to the identity, i.e.  $\mathcal{K}(1)\propto\ID$, we find 
\begin{equation}\label{eq:logderiv2}
-\frac{1}{4}\frac{\partial}{\partial x}\ln T(x)\big|_{x=1}=-\frac{1}{4}\frac{\tr_a \bar{\mathcal{ K}}_a'(1)}{\tr_a \bar{\mathcal{K}}_a(1)}-\frac{1}{2}\frac{\tr_a \bar{\mathcal{K}}_a(1)\mathcal{R}'_{a,1}(1)\mathcal{P}_{a,1}}{\tr_a \bar{\mathcal{K}}_a(1)}-\frac{1}{2}\sum_{k=1}^{N-1}\mathcal{R}'_{k-1,k}(1)\mathcal{P}_{k-1,k}-\frac{1}{4}\frac{{{\mathcal{K}}}_N'(1)}{{\mathcal{K}}_N(1)}\,,
\end{equation} 
see~\cite{Sklyanin:1988yz}. Here we used the symmetry of the R-matrix \eqref{eq:symm}. The first term in \eqref{eq:logderiv2} is proportional to the identity and can be subtracted without spoiling commutativity by setting 
\begin{equation}
 c_0=\frac{1}{4}\frac{\tr_a \bar{\mathcal{ K}}_a'(1)}{\tr_a \bar{\mathcal{K}}_a(1)}
\end{equation} 
in the Hamiltonian \eqref{eq:logderiv1}. The interesting points of the K-matrix that enter the Hamiltonian  are thus $\bar{ \mathcal{K}}(1)=M^{-1} \mathcal{K}(q^{-1})|_{R\to L}$ , $\mathcal{K}(1)$ and $\mathcal{K}'(1)$. 
As we are currently not aware of a generic expression of the K-matrix, we follow a procedure similar to the one proposed by Sutherland to determine the Hamiltonian density \cite{sutherland1970two}. 
More precisely we will study the boundary Yang-Baxter equations in the form given in \eqref{manga1} and \eqref{manga2} for the K-matrices at the points $y=q^{-1}$ and $y=1$.  A similar equation can be obtained for after taking the derivative of these equations also for $\mathcal{K}'(1)$ by differentiating the relations \eqref{manga1} and \eqref{manga2}. We will not attempt to compute the scalar $c_0$ as it is not relevant for our argument.

\subsection{Bulk}\label{sec:bulkT}

At the special point $x=1$ the R-matrix above reduces to the permutation operator such that
\begin{equation} 
 \langle m,n|\mathcal{R}(1)|m',n'\rangle =\delta_{m,n'}\delta_{n,m'}\,,
\end{equation} 
i.e. $\mathcal{R}(1)=\mathcal{P}$. 
This follows from the property $
 \Phi_{q}(i|j;y;y)=\delta_{i,j}$ discussed in \cite{Mangazeev:2019rzf}. 
Using $
 \Phi_{q}(i|j;1;y)=\delta_{i,0}$
 which can also be found in \cite{Mangazeev:2019rzf}  we  obtain  
\begin{equation}
\begin{split}
 \langle m,n|\mathcal{R}'(1)\mathcal{P}|m',n'\rangle =\delta_{m+n,m'+n'} \bigg(&\partial_x\Phi_{q^2}(n'-n|n';x^{-2},x^{-2}q^{4s}) +\partial_x\Phi_{q^2}(m|m';x^2q^{4s},q^{4s})\bigg)\bigg|_{x=1}\,.
\end{split}
\end{equation} 
The two terms exactly lead  the rates of particles moving to the left and to the right. We find 
\begin{equation}\label{eq:hleft}
 \left.\partial_x\Phi_{q^2}(n'-n|n';x^{-2},x^{-2}q^{4s})\right|_{x=1}\delta_{m+n,m'+n'}=-2\alpha_+(n)\delta_{n,n'}\delta_{m,m'}+2 \beta_{+}(n',n'-n)\delta_{m+n,m'+n'}\delta_{n<n'}
\end{equation} 
\begin{equation}\label{eq:hright}
 \left.\partial_x\Phi_{q^2}(m|m';x^2q^{4s},q^{4s})\right|_{x=1}\delta_{m+n,m'+n'}=-2\alpha_-(m)\delta_{n,n'}\delta_{m,m'}+2 \beta_{-}(m',m'-m)\delta_{m+n,m'+n'}\delta_{m<m'}
\end{equation}  
and thus
\begin{equation}
\mathcal{H}=-\frac{1}{2}\mathcal{R}'(1)\mathcal{P}\,.
\end{equation} 
The   two relations \eqref{eq:hleft} and \eqref{eq:hright} are shown in Appendix~\ref{app:hopper}.

We remark that in the construction of the stochastic R-matrix \cite{Kuniba:2016fpi} the rates  arise from special points of the transfer operators leading to either left or right moving particles. Here we consider the full transfer matrix to obtain the natural generalisation to the q-Hahn process with hoppings in both directions from the logarithmic derivative. For the closed chain the q-Hahn Hamiltonian  has been constructed as a superposition   in \cite{Kuniba:2016fpi}.

\subsection{Left boundary}
The goal of this subsection is to obtain $\bar{\mathcal{K}}(1)$ such that the boundary term $B_L$  in \eqref{eq:bd1} arises from  \eqref{eq:logderiv2}. More precisely
\begin{equation}\label{eq:b1term}
B_L= 
 \frac{\tr_a \bar{\mathcal{K}}_a(1)\mathcal{H}_{a,1}}{\tr_a \bar{\mathcal{K}}_a(1)}\,,
\end{equation} 
where 
\begin{equation}
 \bar{\mathcal{K}}(1)=\left.D^{-1}\mathcal{K}(q^{-1})\right|_{R\to L}\,,
\end{equation}
with $D$ given in \eqref{eq:dmat}.
Here we take inspiration from the rational case where it was found that the K-matrix at the special point becomes a rank $1$ matrix.  

Inserting the equations \eqref{manga1} and \eqref{manga2} into a computer algebra system at $y=q^{-1}$ we find that for
\begin{equation}
 t^+_R=-\rho_R\,,\qquad  t^-_R=1\,,\qquad \kappa_R=1\,,\qquad \nu_R=-q^{2s-2}\,,
\end{equation} 
there exists such a degenerate solution
\begin{equation}
 {\mathcal{K}}_{ij}\left(q^{-1}\right)=c_1\left(q^{2} \rho_R \right)^i \frac{  \left(q^{4 s};q^2\right){}_i}{\left(q^2;q^2\right){}_i}\,,
\end{equation} 
that is independent of the column index $j$. Here $c_1$ denotes an overall constant that drops from the final result for the boundary term. It follows  that the other K-matrix can be expressed as
\begin{equation}
 \bar{\mathcal{K}}_{ij}(1)=c_1\,  \rho_L  ^i \frac{  \left(q^{4 s};q^2\right){}_i}{\left(q^2;q^2\right){}_i}\,.
\end{equation} 
The trace  immediately  yields
\begin{equation}\label{eq:trform}
 \tr  \bar{\mathcal{K}}\left(1\right)=c_1\,  \frac{  \left(q^{4 s}\rho_L ;q^2\right){}_\infty}{\left(\rho_L;q^2\right){}_\infty}\,,
\end{equation} 
using the q-analog of the binomial theorem that  relies  on the conditions $|\rho_L|<1$ and $|q^2|<1$, see \cite[(1.3.2)]{gasper2004basic}.
We can now compute the expression in \eqref{eq:b1term} using the explicit form of the Hamiltonian density obtained in the previous Section~\ref{sec:bulkT}. One finds
\begin{equation}\label{eq:trkh}
\begin{split}
 \frac{\tr_a \left( \bar{\mathcal{K}}_a\left(1\right)\mathcal{H}_{a1}\right)}{ \tr  \bar {\mathcal{K}}\left(1\right)}|m_1\rangle=\left(\alpha_+(m_1) +\sum_{k=1}^\infty\frac{\rho_L^k}{1-\gamma^{k}}\right)|m_1\rangle&
-\sum_{k=1}^{m_1} \beta_+(m_1,k)|m_1-k\rangle\\& -\sum_{k=1}^\infty\frac{\rho_L^k}{1-\gamma^{k}}|m_1+k\rangle \,.
\end{split}
\end{equation} 
The details of the computation are given in Appendix~\ref{app:leftbnd}.

\subsection{Right boundary}\label{sec:rib}

To show that the boundary term $B_R$ arises from the transfer matrix via
\begin{equation}\label{eq:BRR}
 B_R=-\frac{1}{4}\frac{{{\mathcal{K}}}_N'(1)}{{\mathcal{K}}_N(1)}\,,
\end{equation} 
 we differentiate the two relations \eqref{manga1} and \eqref{manga2}.
We impose that the K-matrix is proportional to the identity at the special point $y=1$
\begin{equation}
 \mathcal{K}(1)=\frac{1}{4}\ID\,,
\end{equation} 
which is consistent with  \eqref{manga1} and \eqref{manga2}
and set
\begin{equation}\label{eq:identification}
  t^+_R=-\rho_R\,,\qquad  t^-_R=1\,,\qquad \kappa_R=1\,,\qquad \nu_R=-q^{-2s}\,,\,.
\end{equation}  
Then the equations arising from the boundary Yang-Baxter equation above turn into
\begin{equation}\label{manga1h}
 \begin{split}
  -\rho_R q^{2-4s}\left(1-q^{2(j+2s)}\right)\mathcal{K}'_{j,l+1}(1)
 +q^{-4s}\left(1-q^{2+2l}\right)\mathcal{K}'_{j+1,l}(1)\\
 -q^{2}\left(q^{2j}-q^{2l}\right)\left(1+q^{2-4s}\rho_R\right)\left(\mathcal{K}'_{j+1,l+1}(1)+\frac{1}{2}\delta_{j+1,l+1}\right)\\
 +\rho_R q^{2-4s}\left(1-q^{2(1+l+2s)}\right)\left(\mathcal{K}'_{j+1,l+2}(1)+\delta_{j+1,l+2}\right)\\
 - q^{-4s}\left(1-q^{4+2j}\right)\left(\mathcal{K}'_{j+2,l+1}(1)+\delta_{j+2,l+1}\right)=0\,,
 \end{split}
\end{equation} 
and
\begin{equation}\label{manga2h}
 \begin{split}
 -\rho_R q^{2(2+l)}\left(1-q^{2(j+2s)}\right)\mathcal{K}'_{j,l+1}(1)
 +q^{2(1+j)}\left(1-q^{2+2l}\right)\mathcal{K}'_{j+1,l}(1)\\
 -q^{2}\left(q^{2j}-q^{2l}\right)\left(\rho_R+1\right)\mathcal{K}'_{j+1,l+1}(1)
 +\rho_R q^{2(1+j)}\left(1-q^{2(1+l+2s)}\right)\mathcal{K}'_{j+1,l+2}(1)\\
 - q^{2l}\left(1-q^{4+2j}\right)\mathcal{K}'_{j+2,l+1}(1)=0\,.
 \end{split}
\end{equation} 
 Using a computer algebra system we can verify that 
\begin{equation}\label{eq:trk}
 \mathcal{K}'_{i,j}(1)=-\left(\alpha _-(i)+\sum_{k=1}^\infty\frac{\rho_R^k}{q^{-2k}-1}\right)\delta_{i,j}+\beta_-(j,j-i)\delta_{i< j}+\frac{\rho_R^{i-j}}{q^{-2(i-j)}-1}\delta_{i>j}\,,
\end{equation}  
solves both of the equations \eqref{manga1h} and \eqref{manga2h}. To keep track of the differerent contributions on the diagonal, upper and lower triangular part  in \eqref{eq:trk} we distinguish five cases that are exemplified in Figure~\ref{fig:mat} for which the equations are verified separately.
We thus recover the right boundary in \eqref{eq:bd2} from \eqref{eq:BRR}.

\begin{figure*}[t!]
    \centering
    \begin{minipage}{.19\columnwidth}
        \pfive
        \caption*{$j-l>1\qquad$}
    \end{minipage}
    \begin{minipage}{.19\columnwidth}
      \pfour
        \caption*{$j-l=1\qquad$}
    \end{minipage}
    \begin{minipage}{.19\columnwidth}
    \pthree
        \caption*{$j=l\qquad$}
    \end{minipage}
    \begin{minipage}{.19\columnwidth}
        \ptwo
        \caption*{$l-j=1\qquad$}
    \end{minipage}
    \begin{minipage}{.19\columnwidth}
        \pone
        \caption*{$l-j>1\qquad$}
    \end{minipage}
    \caption{Case distinctions of the relations \eqref{manga1h} and \eqref{manga2h}. The boxes denote the matrix entries of the K-matrix with the diagonal  colored in  grey. The entries affected by the relations \eqref{manga1h} and \eqref{manga2h} are indicated in red color.}
    \label{fig:mat}
\end{figure*}

\section{Totally asymmetric and symmetric case}\label{sec:otasep}
In this subsection we discuss two limiting cases of the derived q-Hahn process with boundaries, i.e. the totally asymmetric  ($\mu\to 0$) and the symmetric ($q\to1$)  case. The latter process has been introduced in \cite{Frassek:2019vjt}. We remark the analogy with the ASEP where the corresponding  limiting cases  lead to the TASEP and the SSEP.

\subsection{Totally asymmetric case: $\mu\to 0$}
The open q-Hahn process \eqref{eq:ham} naturally yields a totally asymmetric process in the limit when $\mu$ tends to zero. The bulk term turns into
\begin{equation}\label{eq:hactb}
\begin{split}
 \mathcal{H}|m\rangle \otimes |m'\rangle=\alpha_-(m) |m\rangle \otimes |m'\rangle-\sum_{k=1}^m \beta_-(m,k) |m-k\rangle \otimes |m'+k\rangle \,,
 \end{split}
\end{equation} 
while the boundaries become 
\begin{equation}\label{eq:bd1b}
\begin{split}
B_L|m\rangle = \sum_{k=1}^\infty\frac{\rho_L^k}{1-\gamma^{k}}|m\rangle
  -\sum_{k=1}^\infty\frac{\rho_L^k}{1-\gamma^{k}}|m+k\rangle \,,
\end{split}
\end{equation} 
and
\begin{equation}\label{eq:bd2b}
\begin{split}
B_R|m\rangle = \alpha_-(m)  |m\rangle
-\sum_{k=1}^m\beta_-(m,k)|m-k\rangle  \,.
\end{split}
\end{equation}  
This Hamiltonian describes a process where particles are inserted at the first site with a rate governed by the parameter $\rho_L$ and are then transported to site $N$ at rate $\beta_-$. At the last site they are extracted at the same rate $\beta_-$.
 
\subsection{Symmetric case: $q\to 1$}
The rational limit of the bulk rates yields the symmetric harmonic process discussed in \cite{Frassek:2019vjt}. Recalling the definition \eqref{eq:qq}, one finds
\begin{equation}\label{eq:1}
 \lim_{q\to 1}\log (q^{-2})\beta_\pm(m,k)=\frac{1}{k}\frac{\Gamma(m+1)\Gamma(m-k+2s)}{\Gamma(m-k+1)\Gamma(m+2s)}
\end{equation} 
and
\begin{equation}\label{eq:2}
 \lim_{q\to 1}\log (q^{-2})\alpha_\pm(m)=\sum_{k=0}^{m-1}\frac{1}{k+2s}\,.
\end{equation} 
These are the bulk rates introduced in \cite{Frassek:2019vjt}, see also \cite{melo}.
For the boundaries we obtain

\begin{equation}\label{eq:bdrat}
\begin{split}
\lim_{q\to1}\log(q^{-2})
B_{L,R}|m\rangle =\left(\sum_{k=0}^{m-1}\frac{1}{k+2s} +\sum_{k=1}^\infty\frac{\rho_L^k}{k}\right)|m\rangle&
-\sum_{k=1}^m \frac{1}{k}\frac{\Gamma(m+1)\Gamma(m-k+2s)}{\Gamma(m-k+1)\Gamma(m+2s)}|m-k\rangle\\& -\sum_{k=1}^\infty\frac{\rho_L^k}{k}|m+k\rangle \,,
\end{split}
\end{equation} 
in agreement with \cite[(2.28)]{Frassek:2019vjt}. The limit above is obtained using   \eqref{eq:1} and \eqref{eq:2} as well as
\begin{equation}
  \lim_{q\to 1}\log (q^{-2})\frac{\rho_L^k}{1-\gamma^{k}}=\frac{\rho_L^k}{k}\,.
\end{equation}  

\section{Conclusion}\label{sec:conc}

The definition of the q-Hahn process with open boundary conditions allows for the study of its microscopic properties along the lines of its prominent finite-dimensional cousin known as the Asymmetric Simple Exclusion Process (ASEP). In the $q\to 1$ limit we recover the symmetric hopping model with boundaries introduced in \cite{Frassek:2019vjt} while $\mu\to0$ leads to an analogy of the open TASEP. It would be interesting to generalise the results obtained so far in the symmetric setting to the asymmetric one. In particular, for the symmetric model the steady state has been computed exactly in \cite{Frassek:2021yxb} and a dual Levy process was obtained in \cite{Franceschini}.

Further, it would be desirable to obtain the full K-matrix for non-compact spin chains  and derive the introduced boundary conditions from it.  
 As in the rational case \cite{Frassek:2019vjt}, the K-matrices for the q-Hahn process  should arise from a universal or algebraic K-operator, that can hopefully be expressed in terms of $U_q(sl_2)$ generators. The derived boundary terms of the Hamiltonian should then be obtained  after specifying the representation and identifying the free parameters as we did in Section~\ref{sec:ham}.  To derive such expression it is promising to explore the relation  between the boundary Yang-Baxter equation and quantum symmetric pairs  to reduce the boundary Yang-Baxter equation to a system of linear equations, see in particular \cite{Delius:2002ma,Baseilhac:2016uxv,Nepomechie:2002js} and the recent work \cite{Appel:2022scp}. In particular, it would be interesting to see whether the intertwining relations and the q-Onsager algebra, see \cite{Terwilliger:2004pt,Baseilhac:2004gi} as well as the discussion in \cite{Tsuboi:2019qac}, simplify for the specific choice of boundary parameters relevant for the stochastic solution that we determined here.  
 As already mentioned in the introduction, such algebraic expressions are advantageous when studying stochastic dualities. We hope to return to the study of the intertwining relations in a future work. 
 We also like to point out that several expressions for K-matrices exist in the literature that may be useful to obtain a representation of the K-matrix of the q-Hahn process. The approach most suitable for our purposes is discussed in \cite{Tsuboi:2019qac} where an algebraic form of the triangular K-matrix was obtained. This algebraic form of the K-matrix can be used to recover the boundary term \eqref{eq:bd2} for the case $\rho_R=0$, cf.~\eqref{eq:bd2b}, when evaluating the logarithmic derivative at the special point, see~\eqref{eq:BRR}, for a certain choice of boundary parameters.~\footnote{The same paper proposes a K-matrix that, according to the author,  does not \textit{seem}
to satisfy all intertwining relations. It would be very interesting to sort out these irregularities.}  Further details will be reported elsewhere.  Other approaches to the construction of K-matrices that may be useful for the derivation of the full K-matrix for non-compact spin chains can be found in   \cite{Delius:2002ad,Lazarescu,Mangazeev:2019rzf,Kuniba:2018qzk,Kuniba:2019ans,Kuniba:2019syl}.
One may also ask whether the boundary conditions we introduced are the most general stochastic boundaries. This question can likely be answered once the full K-matrix is known explicitly. 

Finally  it would be very interesting to study whether our prescription to obtain boundary terms of the integrable Hamiltonian  can be formalised leading to a method that allows bootstrapping integrable Hamiltonians for open chains in analogy to Sutherland's approach.  In particular we are not aware of a reference about the special point where the K-matrix reduces to rank $1$ in the literature.

\paragraph{Acknowledgments}
I like to thank Cristian Giardin\`{a}, Istv\'{a}n M. Sz\'{e}cs\'{e}nyi, Zengo Tsuboi and  especially 
Rodrigo A. Pimenta for inspiration and fruitful discussions. I also thank the referees for their useful comments. Further I thank the organisers and participants of the GGI workshop \textit{``Randomness, Integrability and Universality''}, where part of the work has been carried out, for creating a stimulating environment. I thank the  Galileo Galilei Institute (GGI) for support and hospitality during this scientific program. 
RF has partly been  supported by the German research foundation (DFG)
Research Fellowships Programme 416527151.

\appendix

\changelocaltocdepth{1}
\section{Sum formulas}\label{app:sumrates}
In this appendix we argue that 
\begin{equation}
\alpha_\pm(m)
=\sum_{k=1}^{m} \beta_\pm(m,k) \,.
\end{equation} 
In the case of $\alpha_-$ the sum on the right hand side can be evaluated as follows. First we note that
\begin{equation}\label{eq:sumrates1}
 \begin{split}
 \sum_{k=1}^m
\beta_-(m,k)&=\frac{(\gamma;\gamma)_m}{(\mu ;\gamma)_m}\sum_{k=1}^\infty
 \frac{1 }{1-\gamma^k}
\frac{ (\mu ;\gamma)_{m-k}}{ (\gamma;\gamma)_{m-k} }\\
&= \frac{(\gamma^{-m} ;\gamma)_{\infty}}{(\gamma^{1-m-2s};\gamma)_{\infty}}\sum_{k=1}^\infty
 \frac{1 }{1-\gamma^k}
\frac{ (\gamma^{1+k-m-2s};\gamma)_{\infty}}{ (\gamma^{k-m} ;\gamma)_{\infty} } \gamma^{(1-2s)k}  \,,
 \end{split}
\end{equation} 
where  we used the relation 
\begin{equation}
 \frac{(a;q^2)_{n-k}}{(b;q^2)_{n-k}}= \frac{(a;q^2)_{n}}{(b;q^2)_{n}} \frac{(q^{2-2n}b^{-1};q^2)_{k}}{(q^{2-2n}a^{-1};q^2)_{k}}\left(\frac{b}{a}\right)^k\,,
\end{equation}  
combined with $(a;q)_n=\frac{(a;q)_\infty}{(aq^n;q)_\infty}$
for $a,b\in\mathbb{C}$,
see \cite{gasper2004basic}. Next,  introducing the q-deformed digamma function $\widetilde\psi_\gamma$ following \cite{KRATTEN} we obtain
\begin{equation}\label{eq:sumrates2}
 \begin{split} \sum_{k=1}^m
\beta_-(m,k)&=-\frac{1}{\log\gamma}\left(\widetilde\psi_\gamma( {1-m-2s})-\widetilde\psi_\gamma( {1-2s})\right)
\\&=-\frac{1}{2 \log q}\left(\psi_q( {1-m-2s})- \psi_q( {1-2s})-m\log q\right)\\& =-\frac{1}{2 \log q}\left(\psi_q( {m+2s})- \psi_q( {2s})-m\log q\right)\,.
 \end{split}
\end{equation}  
Here $ \psi_q$ denotes the digamma function used in  \cite{Bytsko:2001uh}. The two definitions are related via
\begin{equation}
\psi_q(x)=\widetilde\psi_{\gamma}(x)+\frac{1}{2} (3-2 x) \log q\,,
\end{equation} 
where $\gamma=q^2$,  for further details see also \cite{Frassek:2019isa} where the same notation is used.
Using the reflection formula for the digamma function, cf.~\cite{Frassek:2019isa}, we recover the expression for $\alpha_-$ given in the same reference.
 The final expression in \eqref{eq:sumrates2} in terms of the digamma functions then leads to the desired sum in \eqref{eq:alphas} when expanding the digamma function in analogy to the rational case, see \cite{KRATTEN}. The case of $\alpha_+$ can be deduced from the case studied  here in detail when substituting $q\to q^{-1}$.

\section{Evaluation of Hamiltonian density}\label{app:hopper}
In this appendix we show the relations in \eqref{eq:hleft} and \eqref{eq:hright} that yield the  Hamiltonian density.

\paragraph{Proof of  \eqref{eq:hleft}.}
To show \eqref{eq:hleft} we first consider  the function
\begin{equation}
 \Phi_{q^2}(m|n;x^{-2},x^{-2}q^{4s})=q^{4ms}
\frac{\left(x^{-2};q^2\right)_{m}\left(q^{4s};q^2\right)_{n-m} }{\left(x^{-2}q^{4s};q^2\right)_{n }}
 \left[\begin{array}{c}
n                                                             \\
m                                                                    \end{array}
\right]_{q^2}\,,
\end{equation} 
with $m\leq n$, cf.~\eqref{eq:phii}.
The function $\Phi_{q^2}(m|n;x^{-2},x^{-2}q^{4s})$ vanishes for negative integer values $m=-1,-2,\ldots$ as the term $(q^2;q^2)_{m}=(q^{2+n'-n};q^2)_{m}^{-1}$ from the q-binomial appears in the denominator. For $0\leq m\leq n$ the derivative of the function above can be written as
\begin{equation}
\begin{split}
\partial_x \Phi_{q^2}(m|n;x^{-2},x^{-2}q^{4s})=q^{4ms}
 \left(q^{4s};q^2\right)_{n-m}
 \left[\begin{array}{c}
n                                                             \\
m                                                                    \end{array}
\right]_{q^2}\Bigg[&\frac{\partial_x\left(x^{-2};q^2\right)_{m} }{\left(x^{-2}q^{4s};q^2\right)_{n }}\\&-\frac{\left(x^{-2};q^2\right)_{m} }{\left(x^{-2}q^{4s};q^2\right)^2_{n }}\partial_x \left(x^{-2}q^{4s};q^2\right)_{n }\Bigg]\,,
\end{split}
\end{equation} 
where the derivative of the q-Pochhammer symbol yields
\begin{equation}\label{eq:der1}
 \frac{\partial}{\partial x}(x;q)_n 
 =-(x;q)_n\sum_{p=0}^{n-1}\frac{q^p}{1-xq^p}\,.
\end{equation} 
We need to take extra care when $x\to1$. Here we obtain
\begin{equation}\label{eq:der2}
\left.
 \frac{\partial}{\partial x}(x;q)_n\right|_{x=1} 
 =-(q;q)_{n-1}\,.
\end{equation} 
This leads the first equality    \eqref{eq:hleft} which we wanted to show
\begin{equation}
\begin{split}
\left.
\partial_x \Phi_{q^2}(m|n;x^{-2},x^{-2}q^{4s})\right|_{x=1}&
=2q^{4ms}
 \left[\begin{array}{c}
n                                                             \\
m                                                                    \end{array}
\right]_{q^2}\frac{
 \left(q^{4s};q^2\right)_{n-m} }{\left(q^{4s};q^2\right)_{n }}\left[\left(q^2;q^2\right)_{m-1} -\delta_{m,0}
  \sum_{k=0}^{n-1}\frac{q^{4s}q^{2k}}{1-q^{4s}q^{2k}}\right]\\
  &
=-2 \left[
  \alpha_+(m)\delta_{m,0}-\beta_{+}(n,m) \right]\,. 
\end{split}
\end{equation}

\paragraph{Proof of  \eqref{eq:hright}.}
The second relation \eqref{eq:hright} involves the function
\begin{equation}
 \Phi_{q^2}(m|n;x^2q^{4s},q^{4s})=x^{-2m}
\frac{\left(x^2q^{4s};q^2\right)_{m}\left(x^{-2};q^2\right)_{n-m} }{\left(q^{4s};q^2\right)_{n}}
 \left[\begin{array}{c}
n                                                  \\
m                                                                  \end{array}
\right]_{q^2}\,,
\end{equation} 
that vanishes for $m>n$ because of the q-binomial. The derivative yields
\begin{equation}\label{eq:aap0}
\begin{split}
 \partial_x\Phi_{q^2}(m|n;x^2q^{4s},q^{4s})&=-2mx^{-1}\Phi_{q^2}(m|n;x^2q^{4s},q^{4s})\\
 &\qquad +
\frac{\partial_x \left(x^2q^{4s};q^2\right)_{m}\left(x^{-2};q^2\right)_{n-m} + \left(x^2q^{4s};q^2\right)_{m}\partial_x\left(x^{-2};q^2\right)_{n-m} }{\left(q^{4s};q^2\right)_{n}}
 \left[\begin{array}{c}
n                                                  \\
m                                                                  \end{array}
\right]_{q^2}\,.
\end{split}
\end{equation} 
The first term contributes a Kronecker delta  for $x\to1$ since $(1,q^2)_{n-m}$ is non-zero only for $n-m= 0$.
Further, for the terms involving a derivative of the Pochhammer symbol  we obtain
\begin{equation}\label{eq:aap1}
  \left[\begin{array}{c}
n                                                  \\
m                                                                  \end{array}
\right]_{q^2}\left.\left(x^{-2};q^2\right)_{n-m}\partial_x \left(x^2q^{4s};q^2\right)_{m}\right|_{x=1}=-2\delta_{m, n}\left(q^{4s};q^2\right)_{m}\sum_{p=0}^{m-1}\frac{q^{2p}}{q^{-4s}-q^{2p}}\,,
\end{equation} 
and
\begin{equation} \label{eq:aap2}
\left.\partial_x\left(x^{-2};q^2\right)_{n-m} \right|_{x=1}= 2
 \left(q^{2};q^2\right)_{n-m-1}\,,
\end{equation} 
cf.~\eqref{eq:der1} and \eqref{eq:der2}. Inserting \eqref{eq:aap1} and \eqref{eq:aap2} into \eqref{eq:aap0} leads us to \eqref{eq:hright} which we wanted to show
\begin{equation}
\begin{split}
\left.
 \partial_x\Phi_{q^2}(m|n;x^2q^{4s},q^{4s})\right|_{x=1}&=-2m\delta_{m,n} -2\alpha_+(m)\delta_{m,n}+2
\frac{ \left(q^{4s};q^2\right)_{m}\left(q^{2};q^2\right)_{n-m-1} }{\left(q^{4s};q^2\right)_{n}}
 \left[\begin{array}{c}
n                                                  \\
m                                                                  \end{array}
\right]_{q^2}\\
&= -2\left[\alpha_-(m)\delta_{m,n}-\beta_-(n,n-m)\right]\,.
\end{split}
\end{equation} 
In this last step we used the relation between the diagonal terms of the Hamiltonian density given by
\begin{equation}
 \alpha_+(m)+m=\sum_{p=0}^{m-1}\left(\frac{q^{2p}}{q^{-4s}-q^{2p}} +1\right)=\sum_{p=0}^{m-1}\frac{q^{-4s}}{q^{-4s}-q^{2p}} =\alpha_-(m)\,.
\end{equation} 

\section{Computations for the left boundary}\label{app:leftbnd}
\paragraph{Proof of  \eqref{eq:trkh}.}
To show that \eqref{eq:trkh} holds we compute its action on   a state of $m_1$ particles while taking the trace in the auxiliary space indicated by the letter $a$. We find
\begin{equation}\label{eq:KHact}
\begin{split}
 \tr_a \bar{ \mathcal{K}}_a\left(1\right)\mathcal{H}_{a1}|m_1\rangle&=c_1\sum_{m_a,n_a=0}^\infty \rho_L^{m_a} \frac{  \left(q^{4 s};q^2\right){}_{m_a}}{\left(q^2;q^2\right){}_{m_a}}\langle n_a|\mathcal{H}_{a1}|m_a,m_1\rangle \\
 &=c_1\sum_{n_a=0}^\infty \rho _L^{n_a} \frac{  \left(q^{4 s};q^2\right){}_{n_a}}{\left(q^2;q^2\right){}_{n_a}}\alpha_-(n_a)|m_1\rangle+ \tr_a  \bar{\mathcal{K}}_a\left(1\right)\alpha_+(m_1)|m_1\rangle\\
 &\qquad- c_1\sum_{m_a>n_a}\rho_L^{m_a} \frac{  \left(q^{4 s};q^2\right){}_{m_a}}{\left(q^2;q^2\right){}_{m_a}}\beta_{-}(m_a,m_a-n_a)|m_1+m_a-n_a\rangle\\
 &\qquad- c_1\sum_{m_a<n_a} \rho_L^{m_a} \frac{  \left(q^{4 s};q^2\right){}_{m_a}}{\left(q^2;q^2\right){}_{m_a}}\beta_{+}(m_1,n_a-m_a)|m_1+m_a-n_a\rangle\,.
 \end{split}
\end{equation} 
The individual terms are evaluated in analogy to \eqref{eq:trform}  in the following:
\paragraph{Diagonal term}
The series in the  first term above can be simplified and written as
\begin{equation}  
\begin{split}
c_1
 \sum_{m=0}^\infty \rho_L ^{m} \frac{  \left(q^{4 s};q^2\right){}_{m}}{\left(q^2;q^2\right){}_{m}}\alpha_-(m)
&=c_1\sum_{m=1}^\infty\rho_L  ^{m} \frac{(q^{4s};q^2)_m  }{  (q^{2} ;q^2)_m}\sum_{k=1}^{m}\,  \beta_-(m,k)\\
&=c_1\sum_{m=1}^\infty\sum_{k=1}^{m}\,  \rho_L  ^{m} \frac{  \left(q^{4 s};q^2\right){}_{m-k}}{\left(q^2;q^2\right){}_{m-k}}\frac{1}{1-q^{2k}}\\
 & =\tr_a  \bar{\mathcal{K}}_a\left(1\right)\sum_{k=1}^\infty\frac{\rho_L^k }{1-q^{2k}}\,.
 \end{split}
\end{equation} 

\paragraph{Insertion term}
Changing variables 
$n_a=m_a-k$, the series in \eqref{eq:KHact} involving $\beta_-$ yields 
\begin{equation}
\begin{split} 
&c_1
 \sum_{n_a=0}^{\infty}\sum_{m_a=n_a+1}^\infty\rho_L^{m_a} \frac{  \left(q^{4 s};q^2\right){}_{m_a}}{\left(q^2;q^2\right){}_{m_a}}\beta_{-}(m_a,m_a-n_a)|m_1+m_a-n_a\rangle\\
 &=c_1
 \sum_{n_a=0}^{\infty}\sum_{k=1}^\infty\rho_L^{k+n_a} \frac{  \left(q^{4 s};q^2\right){}_{k+n_a}}{\left(q^2;q^2\right){}_{k+n_a}}\beta_{-}(k+n_a,k)|m_1+k\rangle 
 \\&
 =\tr_a  \bar{\mathcal{K}}_a\left(1\right)
  \sum_{k=1}^{ \infty}\frac{\rho_L^k }{1-q^{2k}}|m_1+k\rangle\,.
 \end{split}
\end{equation}  
\paragraph{Extraction term}
Similarly, after setting $n_a=m_a+k$,  the series involving $\beta_+$ gives
\begin{equation}
\begin{split}
&c_1
 \sum_{m_a=0}^\infty\sum_{n_a=m_a+1}^\infty \rho_L^{m_a} \frac{  \left(q^{4 s};q^2\right){}_{m_a}}{\left(q^2;q^2\right){}_{m_a}}\beta_{+}(m_1,n_a-m_a)|m_1+m_a-n_a\rangle\\ 
&=c_1
 \sum_{m_a=0}^\infty\sum_{k=1}^{m_1} \rho_L^{m_a} \frac{  \left(q^{4 s};q^2\right){}_{m_a}}{\left(q^2;q^2\right){}_{m_a}}\beta_{+}(m_1,k)|m_1-k\rangle
 \\&
 =\tr_a  \bar{\mathcal{K}}_a\left(1\right)
 \sum_{k=1}^{m_1} \beta_{+}(m_1,k)|m_1-k\rangle\,.
\end{split}
\end{equation} 
 Finally we recover \eqref{eq:b1term}. 
%



{
{
\small
\bibliographystyle{utphys2}
\providecommand{\href}[2]{#2}\begingroup\raggedright\endgroup

}

\noindent\rule{6cm}{0.4pt}

\texttt{Contact: rfrassek@unimore.it}

\end{document}